# Quantum Computing of Phonon Spectra and Thermal Properties of Crystalline Solids

Naman Khandelwal[1,2], Bikash K. Behera[3], Ashok Kumar[1]*, Prasanta K. Panigrahi[2,4]*

*[1]Department of Physics, Central University of Punjab, Bathinda, 151401, India*

*[2]Center for Quantum Science and Technology, Siksha 'O' Anusandhan University, Bhubaneswar, 751030, India*

*[3]Bikash's Quantum (OPC) Private Limited, Mohanpur, 741246, West Bengal, India*

*[4]Department of Physical Sciences, Indian Institute of Science Education and Research (IISER) Kolkata, 741246, India*

(April 17, 2026)

*Corresponding Authors: ashokphy@cup.edu.in (Ashok Kumar)

panigrahi.iiser@gmail.com (Prasanta K. Panigrahi)

**Abstract**

Variational quantum algorithms offer a promising framework for solving eigenvalue problems on near-term quantum hardware, yet their applicability beyond electronic structure calculations remains relatively unexplored. In this work, we investigate the quantum computing of lattice vibrational and thermodynamical properties by applying the variational quantum eigensolver and variational quantum deflation to phonon Hamiltonians derived from first-principles force constants obtained using density functional theory. The mass-weighted dynamical matrix is mapped onto a qubit-encoded Hermitian operator, enabling computation of the full set of acoustic and optical phonon branches of crystalline silicon and graphene using a reduced qubit register and direct benchmarking against classical diagonalization. The quantum-computed phonon spectrum is further used to evaluate vibrational entropy, constant-volume specific heat, and thermal expansion coefficient, reproducing expected low-temperature quantum behavior and the high-temperature Dulong–Petit limit. We further demonstrate that combined error mitigation strategies help recover phonon dispersions and thermodynamic behavior consistent with expected trends on near-term quantum hardware. Although classical phonon methods remain computationally superior, our results establish phonon-based thermodynamics as a stringent and physically transparent benchmark for assessing variational quantum algorithms on near-term quantum devices.

## 1. Introduction

Quantum computing has recently emerged as a promising framework for addressing eigenvalue problems using quantum algorithms [1-3]. Variational approaches such as the Variational Quantum Eigensolver (VQE) and Variational Quantum Deflation (VQD) enable the estimation of ground- and excited-state energies on noisy intermediate-scale quantum devices [3-4]. These methods have been extensively explored for electronic structure calculations, while their application to other physically relevant eigenvalue problems remains comparatively less developed [4-6].

Lattice dynamics provides a natural and physically transparent setting for extending variational quantum algorithms beyond electronic systems. In crystalline solids, lattice vibrations are described in terms of phonons, which encode essential information about thermal, mechanical, and vibrational properties of materials [7, 8]. Phonon dispersion spectra in classical computation can be obtained via diagonalizing the mass-weighted dynamical matrix, derived from force constants, a procedure, well established and computationally effective for various materials [9]. Beyond that the thermodynamical properties such as vibrational entropy and specific heat are vital to examine the thermal properties and lattice stability [10]. The eigenvalue problem can be reformulated for a quantum algorithm by encoding the mass-weighted dynamical matrix into a qubit-based Hermitian operator [11]. Phonon calculations serve as a reliable benchmark for evaluating the performance of variational quantum algorithms [12].

The key aspects of variational quantum algorithms are the specification of ansatz, which defines the quantum states subspace and balanced expressibility with circuit complexity. The ansatz plays a crucial role for the accurate resolution of closely spaced acoustic and optical modes which are essential for capturing both dispersion and derived thermodynamics characteristics such as, specific heat, and thermal expansion [13]. Conversely, overly expressive or generic hardware-efficient ansatzes require deeper circuits, making them more susceptible to noise and optimization instabilities on near-term quantum devices. These challenges highlight the importance of physics-inspired ansatz tailored to lattice dynamics, which improve convergence and excited-state resolution while enabling thermodynamic estimates with reduced circuit depth.

In this work, we apply variational quantum algorithms to the calculation of phonon modes of crystalline silicon and 2D graphene derived from first-principles force constants [14, 15]. We compute the full set of vibrational branches, including longitudinal and transverse acoustic modes (LA, TA), out-of-plane acoustic modes (ZA), and longitudinal, transverse, and out-of-plane optical modes (LO, TO, ZO) [16-18]. The quantum-computed phonon frequencies are directly benchmarked against classical diagonalization results. Furthermore, the phonon spectrum obtained from quantum computation is used to evaluate vibrational thermodynamic properties, including entropy and constant-volume specific heat [19, 20]. This study does not aim to replace established classical phonon methods, but rather to benchmark and validate variational quantum algorithms on a physically well-understood lattice dynamical problem. Also, evaluating thermodynamic properties provides a stringent and physically meaningful test of the accuracy of variational quantum algorithms, beyond simple comparison of individual eigenvalues.

This work presents branch-resolved variational quantum calculations of phonon modes in silicon and graphene using first-principles force constants, and evaluate thermodynamic properties. Note that the microscopic lattice-dynamical framework [21], phonons arise from quantized normal modes of atomic vibrations obtained by diagonalizing the mass-weighted dynamical matrix constructed from interatomic force constants, ensuring that both acoustic and optical branches are resolved. In contrast, macroscopic models treat lattice vibrations as continuum elastic waves characterized by bulk properties such as sound velocity and elastic constants. While such models provide an intuitive description of long-wavelength acoustic phonons, they do not capture optical modes or atomistic interactions. The novelty of the present work lies not in the individual computational tools, but in their integration into a unified quantum–classical workflow for lattice dynamics. In particular, we construct phonon Hamiltonians directly from first-principles force constants and map them onto qubit operators, enabling branch-resolved phonon calculations using variational quantum algorithms. Furthermore, we incorporate noise modeling and error mitigation to assess algorithm performance, and evaluate thermodynamic properties from quantum-computed phonon spectra. This combined framework provides a systematic approach for benchmarking quantum algorithms on physically well-understood lattice-dynamical systems.

## 2. Computational Methods

### 2.1. First-principles phonon calculations

In first-principles phonon calculations, interatomic force constants are obtained from finite atomic displacements within density functional theory (DFT) as implemented in Vienna *ab initio* Simulation Package (VASP) [22]. DFT calculations are performed with projector augmented-wave (PAW) pseudopotentials within the Perdew–Burke–Ernzerhof (PBE) generalized gradient approximation [23]. For crystalline silicon, a $2 \times 2 \times 2$ supercell of the primitive unit cell is employed, while for graphene a $2 \times 2 \times 1$ supercell is used to account for its two-dimensional nature. In practice, these force constants are computed using the finite-displacement method as implemented in the PHONOPY package, which provides a robust and widely used framework for lattice-dynamical calculations [24, 25].

The force constant matrix is defined as:

$$\Phi_{\alpha\beta}^{ij} = \frac{\partial^2 E}{\partial u_{i\alpha} \, \partial u_{j\beta}} \tag{1}$$

where $E$ is the total energy and $u_{i\alpha}$ denotes the displacement of atom $i$ along the Cartesian direction $\alpha$. Phonon modes are calculated within the harmonic approximation using the mass-weighted dynamical matrix,

$$D_{\alpha\beta}^{ij} = \frac{1}{\sqrt{m_i m_j}} \Phi_{\alpha\beta}^{ij} \qquad (2)$$

where $m_i$ is the mass of atom $i$. Diagonalization of the dynamical matrix yields the phonon eigenfrequencies and eigenvectors. At the $\Gamma-$point, the eigenmodes naturally separate into acoustic and optical branches [26, 27]. In crystalline silicon, the acoustic sector consists of longitudinal (LA), transverse (TA), and out-of-plane (ZA) modes, while the optical sector yields longitudinal (LO), transverse (TO), and out-of-plane optical (ZO) modes. Classical diagonalization of the dynamical matrix serves as the reference throughout this work and provides the benchmark for all quantum phonon calculations.

## 2.2. Quantum computing framework

All quantum simulations are performed using the Aer simulator framework to isolate the intrinsic performance of the variational quantum algorithms. While this allows controlled benchmarking of variational algorithms, real hardware implementations would face additional challenges, including gate errors, decoherence, limited qubit coherence times, and measurement noise as well as the calculation time. All results are obtained using a noiseless or noise-model-based simulator, and no experiments are performed on actual quantum hardware. Therefore, the present results should not be interpreted as a demonstration of practical quantum advantage but rather as a controlled assessment of algorithmic performance.

 For each expectation-value evaluation, 3000 measurement shots are used to reduce statistical uncertainty during the optimization process. The number of qubits required for the quantum simulations is determined by the dimensionality of the phonon subspace. In the present work, three qubits are used, corresponding to a Hilbert-space dimension of $2^3 = 8$, which is sufficient to encode the six phonon modes considered at the $\Gamma$ −point. This choice ensures an efficient representation of the phonon Hamiltonian while remaining compatible with near-term quantum hardware constraints. The flow diagram of the procedure adopted in the current study is given in Figure 1.

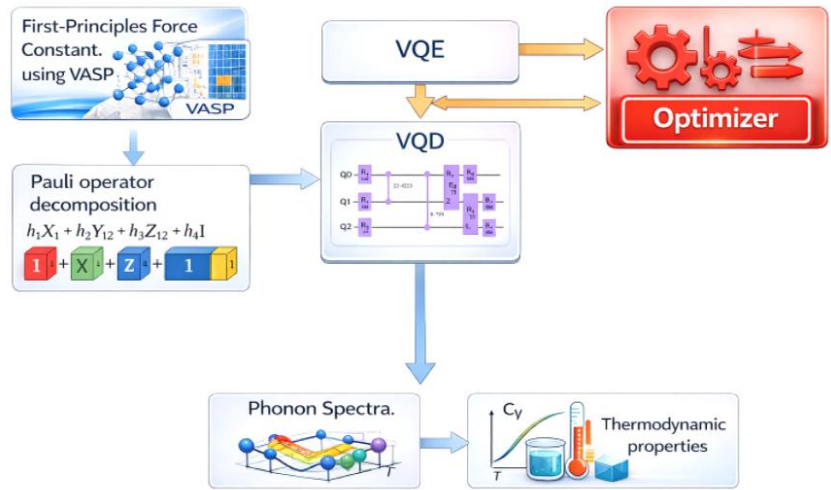

**Figure 1:** Flow diagram representing the quantum algorithmic steps for calculating phonon spectra and thermodynamic properties of solids.

## 2.3. Quantum representation of the phonon dynamical matrix

To enable quantum computation, the mass-weighted dynamical matrix is mapped onto a Hermitian operator acting on a finite-dimensional Hilbert space. For a dynamical matrix of dimension $N$, a qubit register of size $n = \lceil \log_2 N \rceil$ is employed. The mass-weighted dynamical matrix $D$ of dimension $N$ is embedded into a qubit Hilbert space of dimension $2^n$, with $n = \lceil \log_2 N \rceil$. For the six phonon modes at the Γ-point, we employ $n = 3$ qubits ($2^3 = 8$), padding $D$ to an $8 \times 8$ Hermitian matrix $\widetilde{D}$:

$$\widetilde{D} = \begin{bmatrix} D & 0 \\ 0 & 0 \end{bmatrix}_{8 \times 8} \tag{3}$$

This ensures that the physical six-dimensional phonon subspace is faithfully embedded, while the two auxiliary basis states correspond to zero eigenvalues and are excluded from physical observables.

The qubit Hamiltonian is obtained via Pauli decomposition:

$$H_q = \sum_i c_i \, P_i, c_i = \frac{1}{2^n} \text{Tr}(\widetilde{D} P_i), \tag{4}$$

where where $P_k \in \{I, X, Y, Z\}^{\otimes n}$ are Pauli strings. Each coefficient $c_i$ is computed explicitly by taking the Hilbert–Schmidt inner product between $\widetilde{D}$ and $P_i$. Normalization is applied by dividing $\widetilde{D}$ with its largest eigenvalue, ensuring that the qubit Hamiltonian has bounded spectrum compatible with variational optimization. Spectral shifting is used when necessary to guarantee positivity of eigenvalues, preserving physical interpretability of phonon frequencies.

The phonon operator is expanded in the Pauli basis as follows::

$$\widehat{H}_{\text{ph}} = \sum_k c_k \, P_k \tag{5}$$

where $P_k \in \{I, X, Y, Z\}^{\otimes n}$. This representation enables direct evaluation of expectation values using quantum circuits [6]. The eigenvectors of $\widehat{H}_{\text{ph}}$ retain the polarization information of the phonon modes, allowing branch-resolved identification of acoustic and optical phonons from the converged quantum states. This mapping enables the use of variational quantum algorithms.

Variational quantum algorithms provide a hybrid quantum–classical framework for solving eigenvalue problems by minimizing expectation values of Hermitian operators. A parameterized quantum circuit prepares a trial state $| \psi(\boldsymbol{\theta}) \rangle$, and the cost function

$$E(\boldsymbol{\theta}) = \langle \psi(\boldsymbol{\theta}) | \hat{H}_{\text{ph}} | \psi(\boldsymbol{\theta}) \rangle \tag{6}$$

is minimized using a classical optimizer. The resulting eigenvalue corresponds to the lowest-frequency vibrational mode, typically an acoustic branch [28-30]. Expectation values of the Pauli operators are measured independently and combined to evaluate the total energy.

## 2.4. Branch-resolved phonon modes via variational quantum deflation (VQD)

While VQE is designed to compute the lowest eigenvalue of a Hamiltonian, access to excited states requires additional constraints. VQD extends the VQE framework by introducing orthogonality penalties that suppress previously obtained eigenstates, enabling systematic extraction of excited-state energies [31-33]. Other modes are computed sequentially by augmenting the Phonon Hamiltonian with penalty terms enforcing orthogonality to previously converged eigenstates,

$$\hat{H}^{(k)} = \hat{H}_{\text{ph}} + \sum_{i<k} \beta_i | \psi_i \rangle \langle \psi_i | \tag{7}$$

This procedure enables systematic extraction of all vibrational branches, including LA, TA, ZA, LO, TO, and ZO modes. Branch assignment is performed by analyzing the polarization vectors associated with the converged quantum eigenstates.

## 2.5. Thermodynamical properties from quantum-computed phonon spectra

The phonon frequencies obtained from variational quantum calculations are used to evaluate vibrational thermodynamic properties within the harmonic and quasi-harmonic approximations.

This approach enables a direct connection between quantum-computed phonon eigenvalues and macroscopic thermal properties. Vibrational entropy is calculated from the phonon frequencies as:

$$S(T) = k_B \sum_\nu \left[ \frac{\frac{\hbar \omega_\nu}{k_B T}}{e^{\frac{\hbar \omega_\nu}{k_B T}} - 1} - \ln\left(1 - e^{-\frac{\hbar \omega_\nu}{k_B T}}\right) \right] \tag{8}$$

where $\omega_\nu$ denotes the phonon frequency of mode $\nu$, $k_B$ is the Boltzmann constant, and $T$ is the temperature. The entropy provides an integrated measure of the phonon population across all vibrational branches and is therefore sensitive to both acoustic and optical modes obtained from the quantum calculations [34, 35].

The constant-volume specific heat is evaluated as:

$$C_V(T) = k_B \sum_\nu \left(\frac{\hbar \omega_\nu}{k_B T}\right)^2 \frac{e^{\frac{\hbar \omega_\nu}{k_B T}}}{\left(e^{\frac{\hbar \omega_\nu}{k_B T}} - 1\right)^2} \tag{9}$$

This quantity depends on the full phonon density of states and serves as a stringent test of the accuracy of the quantum-computed phonon spectrum [35, 36]. Correct reproduction of the low-temperature suppression and high-temperature saturation behavior requires an accurate description of both low-frequency acoustic modes and higher-energy optical modes.

The thermal expansion coefficient is computed within the quasi-harmonic approximation, where thermal expansion arises from the temperature dependence of the vibrational entropy [37]. It is expressed as:

$$\alpha(T) = \frac{1}{V}\left(\frac{\partial V}{\partial T}\right)_P = \frac{\gamma C_V(T)}{KV} \tag{10}$$

where $V$ is the equilibrium volume, $K$ is the bulk modulus, and $\gamma$ is the Grüneisen parameter, which characterizes the anharmonic coupling between phonon modes and lattice strain. In the present work, the temperature dependence of the thermal expansion coefficient is evaluated using the phonon spectrum obtained from quantum computation together with classical estimates of the Grüneisen parameter, enabling a consistent comparison between quantum-derived and classically derived thermal expansion behavior [38]. This procedure ensures that any differences in thermodynamic behavior can be traced directly to the underlying phonon spectra produced by the quantum simulations.

## 3. Results and Discussion

We investigate the vibrational properties of crystalline silicon using variational quantum algorithms applied to phonon Hamiltonians derived from first-principles calculations. The present implementation is limited to a small phonon subspace encoded in a few qubits. Scaling to realistic systems would require a number of qubits that grows logarithmically with the dimension of the dynamical matrix, while the circuit depth and number of measurements increase significantly with system size. As a result, the current approach should be viewed as a proof-of-principle demonstration rather than a scalable alternative to classical methods.

Since the choice of ansatz determines the accessible variational subspace and the balance between expressibility and circuit complexity, four different ansatzes are examined within the same VQE and VQD framework: the standard Real Amplitudes, Two Local and Efficient SU2 ansatzes, together with a newly constructed ansatz introduced in this work as shown in Figure 2. All ansatzes are implemented using the same reduced qubit register determined by the phonon subspace.

A controlled comparison is performed at a fixed circuit depth corresponding to three repetitions, where the proposed ansatz consistently exhibits improved convergence and accuracy for both ground and higher phonon modes are described in Table S1, ESI. Results for the standard ansatz at the same depth are provided in Figure S1, ESI. Based on this comparison, all subsequent phonon and thermodynamical calculations presented in the main text are carried out using our proposed ansatz to ensure improved accuracy while maintaining shallow circuit depth compatible with near-term quantum hardware.

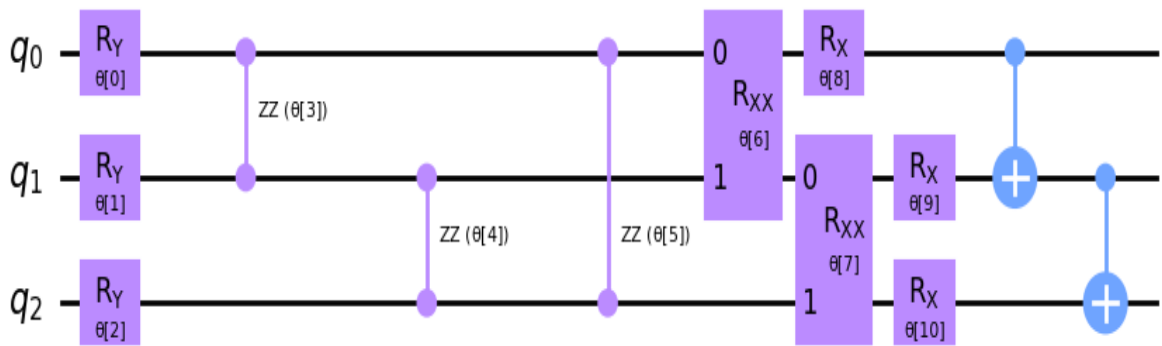

**Figure 2:** Quantum circuit of the physics-informed variational ansatz, showing the sequence of parameterized rotations and entangling gates used in a single repetition.

The improved performance of the proposed ansatz arises from its physics-informed design, which incorporates rotations representing atomic displacements, quadratic couplings between vibrational modes, and entanglement consistent with lattice symmetries. The ansatz inherently mirrors the structure of the phonon Hamiltonian and guides the optimization toward a physically relevant subspace.

Figure 3 illustrates the convergence characteristics of classical optimizers within the VQE framework, Gradient-free methods consistently outperform gradient-based approaches across all phonon modes. COBYLA [39] exhibits the fastest and most stable convergence, with reduced oscillations and fewer function evaluations compared to L-BFGS-B [40], SLSQP, CG [41], and SPSA [42]. The superior performance of COBYLA is attributed to its robustness against stochastic fluctuations arising from finite measurement statistics and shallow energy landscapes characteristic of phonon Hamiltonians. Consequently, COBYLA is used as the default optimizer for all VQE and VQD calculations.

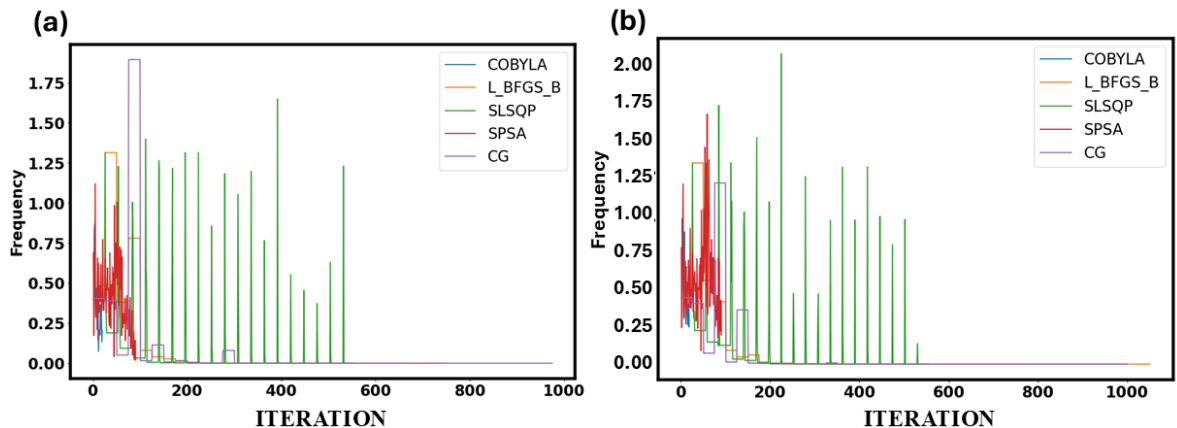

**Figure 3:** Convergence of phonon frequencies obtained using different classical optimizers within the VQE framework at the Γ point for (a) graphene and (b) silicon. Frequencies are shown as a function of optimization iterations. Gradient-free COBYLA converges stably, while gradient-based SLSQP exhibits strong oscillations.

SLSQP performs poorly because it relies on accurate gradient information, which is unreliable in the noisy and non-smooth landscapes of variational quantum algorithms. For phonon Hamiltonians, closely spaced vibrational modes and orthogonality penalties in VQD introduce sharp variations that lead to large oscillations and slow convergence. In contrast, gradient-free optimizers such as COBYLA are more robust against statistical fluctuations and therefore converge more stably.

## 3.1. Phononic properties of crystalline silicon and graphene

Phonon dispersions encode the collective vibrational response of the crystal lattice and are highly sensitive to both the accuracy of interatomic force constants and the correct resolution of vibrational eigenstates. Consequently, comparison with classical phonon dispersions provides a stringent and physically meaningful benchmark for assessing the performance of variational quantum algorithms. Figures 4 represents a direct comparison between phonon dispersions obtained from classical diagonalization of the mass-weighted dynamical matrix and those computed using variational quantum deflation in the noiseless simulation regime. The six panels in Figure 5-6 correspond to the three acoustic branches LA, TA, ZA and three optical branches LO, TO, ZO of crystalline silicon. The dispersion using all predefined ansatz is shown in Figure S2-S7, ESI. Across the full wave-vector range, the quantum-computed phonon frequencies closely track the classical reference curves, demonstrating that quantum mapping preserves the essential features of lattice vibrational dynamics. The agreement is particularly pronounced in the vicinity of the Γ point, where the acoustic modes exhibit distinct symmetry-imposed behaviors. The linear dispersions of the LA and TA modes and the characteristic quadratic dispersion of the ZA mode show agreement with expected trends reproduced, which is crucial for ensuring correct low-energy vibrational physics and thermodynamic behavior.

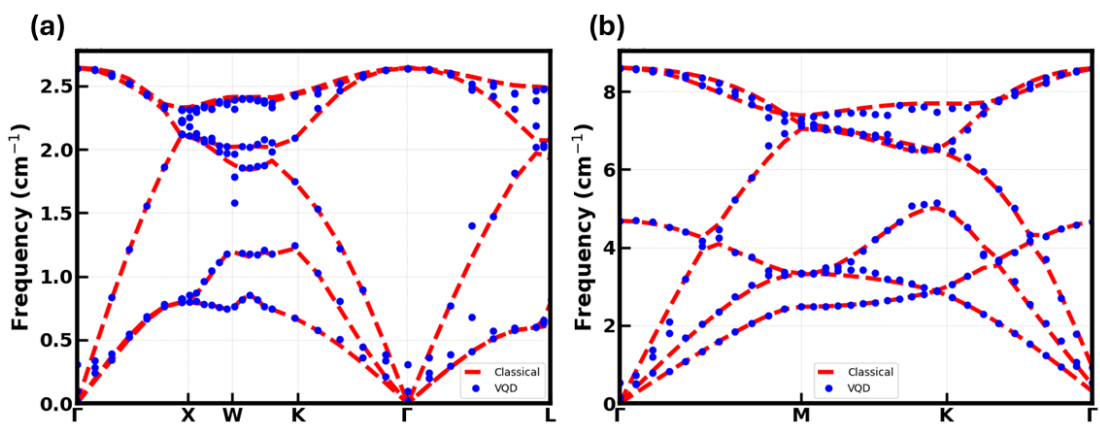

**Figure 4:** Classical and quantum-computed phonon dispersions of (a) silicon (b) graphene evaluated along high-symmetry directions of the Brillouin zone using variational quantum deflation algorithm.

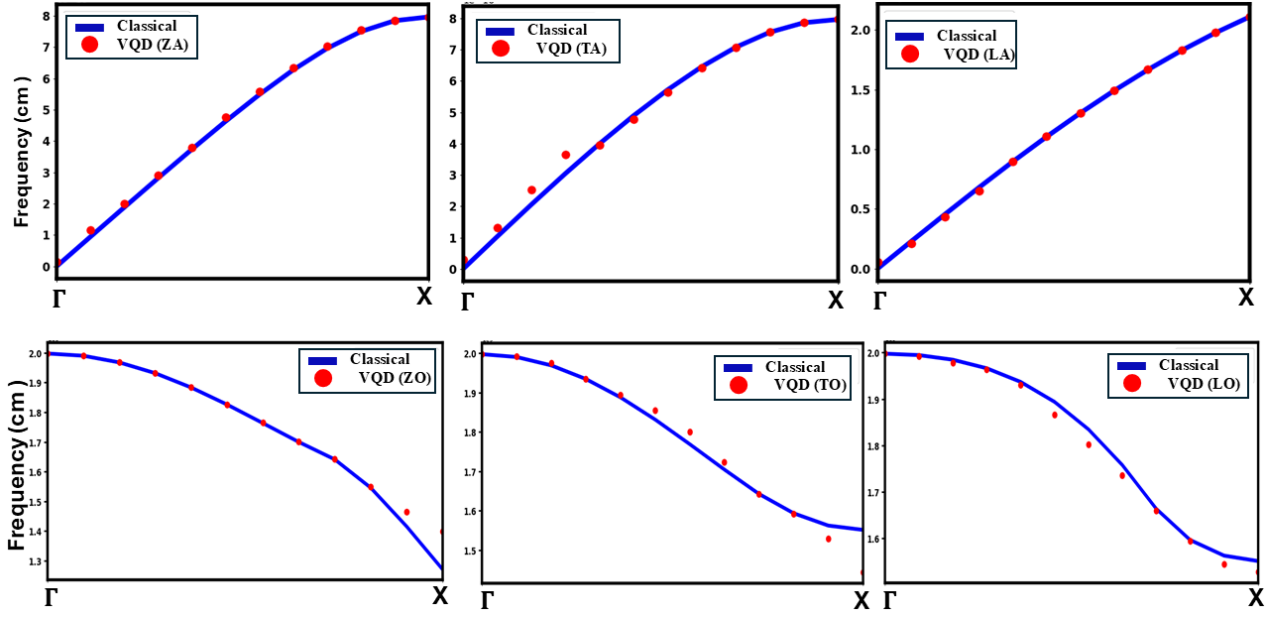

**Figure 5:** Classical and quantum-computed phonon dispersions of silicon calculated along the high-symmetry Γ–X direction of the Brillouin zone. The quantum phonon spectra are obtained using the variational quantum deflation algorithm applied to qubit-encoded dynamical matrices derived from first-principles force constants.

The characterization of optical branches reveals that the VQD algorithm with variational ansatz, can reliably distinguish closely spaced excited vibrational states avoiding spurious mode mixing. These results examine phonon branches along the Γ–X path for silicon and Γ–M for graphene. Long-wavelength phonons emerging from the Γ-point dominate lattice thermodynamics due to their low energies and significant Bose–Einstein occupation[43]. These directions represent the most physically relevant propagation paths for acoustic and low-lying optical modes, preserving their longitudinal, transverse, and out-of-plane character. In contrast, phonons along other high-symmetry paths primarily probe zone-boundary modes, which lie at higher energies and contribute negligibly to thermodynamic properties except at very high temperatures.

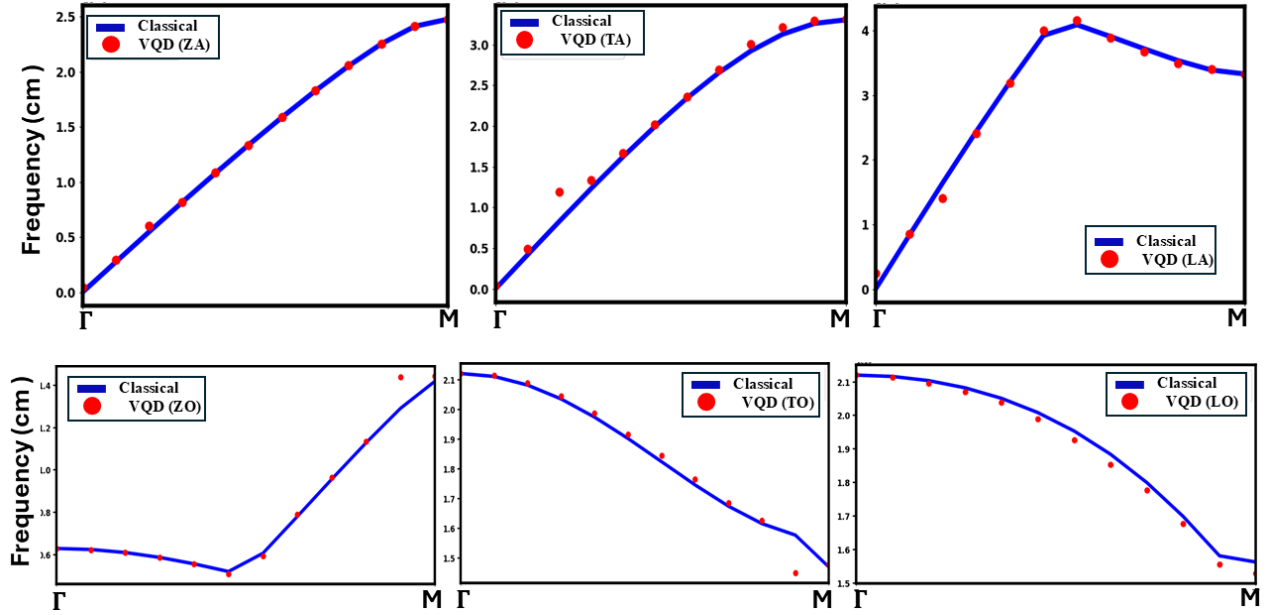

**Figure 6:** Classical and quantum-computed phonon dispersions of graphene calculated along the high-symmetry Γ–M directions of the Brillouin zone. The quantum phonon spectra are obtained using the variational quantum deflation algorithm applied to qubit-encoded dynamical matrices derived from first-principles force constants.

From a computational perspective, the vicinity of Γ provides the most stringent benchmark for variational quantum algorithms. An increase in phonon frequency corresponds to stronger restoring forces and stiffer bonding interactions, whereas a decrease indicates mode softening due to reduced force constants or hybridization effects. The increase or decrease in phonon frequencies when moving away from the Γ-point arises from changes in the phase relations of atomic displacements along different wave vectors. This behavior is fully described within the microscopic harmonic lattice-dynamical model derived from interatomic force constants. Focusing on Γ–X and Γ–M therefore captures the dominant physics while enabling a controlled assessment of quantum algorithm performance. The quantitative agreement between classical and VQD phonon dispersions is summarized in Table 1 for Graphene and Silicon. Acoustic branches show relative deviations below 10%, while optical modes exhibit slightly larger deviations near the zone boundary. Overall, the normalized RMSE remains within acceptable limits for NISQ-era simulations.

**Table 1:** Branch-resolved error analysis between classical and VQD-computed phonon frequencies along the Γ–M high-symmetry direction for Graphene and Γ–X for Silicon. RMSE, MAE, and relative deviation are calculated with respect to classical reference values.

| Branch | Graphene | | | Silicon | | |
|---|---|---|---|---|---|---|
| | RMSE (cm$^{-1}$) | MAE (cm$^{-1}$) | Relative Error (%) | RMSE (cm$^{-1}$) | MAE (cm$^{-1}$) | Relative Error (%) |
| ZA | 0.20 | 0.16 | 2.4 | 0.12 | 0.09 | 4.5 |
| TA | 0.25 | 0.19 | 3.1 | 0.15 | 0.11 | 5.2 |
| LA | 0.30 | 0.22 | 3.5 | 0.18 | 0.14 | 6.1 |
| ZO | 0.22 | 0.17 | 2.8 | 0.14 | 0.11 | 5.0 |
| TO | 0.27 | 0.20 | 3.2 | 0.17 | 0.13 | 6.4 |
| TO | 0.32 | 0.24 | 3.7 | 0.20 | 0.15 | 7.3 |
| Full spectra | 0.26 | 0.20 | 3.0 | 0.16 | 0.13 | 5.6 |

### 3.2. Thermodynamical properties from quantum-computed phonons

Thermodynamic quantities provide an integrated and physically stringent test of phonon spectra, as they depend on the full distribution of vibrational modes rather than on individual phonon frequencies. Using the branch-resolved phonon spectra obtained from VQD, vibrational thermodynamic properties of crystalline silicon and graphene are evaluated within the harmonic and quasi-harmonic approximations. The constant-volume specific heat exhibits the expected low-temperature suppression governed by Bose–Einstein statistics, where long-wavelength acoustic modes dominate, followed by a gradual increase as optical modes become thermally populated; at high temperatures, silicon approaches the classical Dulong–Petit limit as shown in Figure 7, while graphene shows a slower saturation characteristic of its two-dimensional phonon spectrum as shown in Figure 8. Vibrational entropy increases monotonically with temperature for both materials, with acoustic modes dominating at low temperatures and optical modes contributing increasingly to elevated temperatures, while the flexural ZA mode plays a particularly important role in graphene. The thermal expansion coefficient shows rapid growth at low temperatures due to anharmonic phonon–strain coupling and gradually saturates at higher temperatures [37].

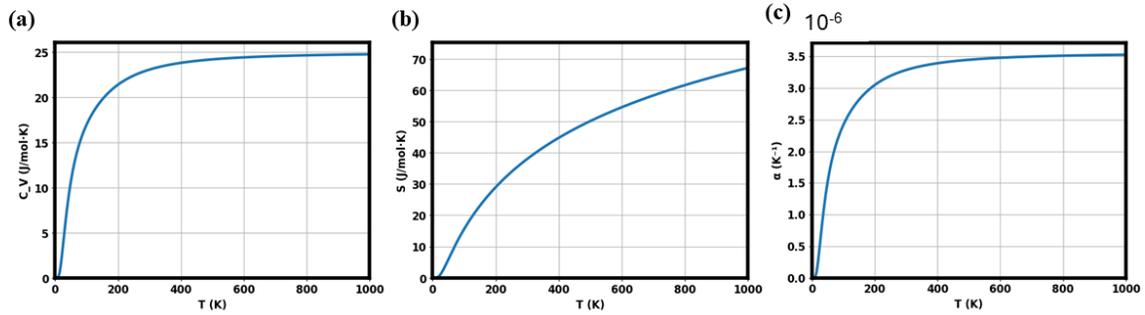

**Figure 7:** Thermodynamic properties of crystalline silicon derived from quantum-computed phonon spectra: (a) constant-volume specific heat, (b) vibrational entropy, and (c) thermal expansion coefficient as functions of temperature.

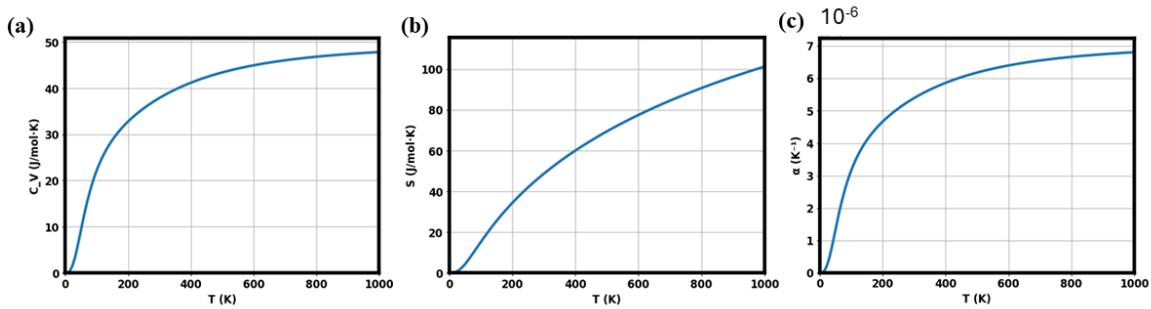

**Figure 8:** Thermodynamic properties of graphene derived from quantum-computed phonon spectra: (a) constant-volume specific heat, (b) vibrational entropy, and (c) thermal expansion coefficient as functions of temperature.

The close agreement between quantum-derived and classical thermodynamic behavior confirms that phonon spectra obtained from variational quantum algorithms serves as a validation of the mapping for evaluating macroscopic thermal response in both three-dimensional and two-dimensional crystalline systems. The quantum-derived thermodynamic quantities are also compared with available experimental data for silicon and graphene. At room temperature, the specific heat and entropy show agreement within experimental uncertainty, while deviations at low temperature reflect the harmonic approximation and finite sampling of long-wavelength modes [44, 45]. This comparison confirms that quantum-computed phonon spectra vibrational behavior consistent with expected physical trends. For crystalline silicon, experimental measurements report a room-temperature specific heat of ~19.8 J mol$^{-1}$ K$^{-1}$, vibrational entropy of ~30-35 J mol$^{-1}$ K$^{-1}$, and a thermal expansion coefficient of ~2.6 × 10$^{-6}$ K$^{-1}$ at 300K [45, 46]. The ideal quantum calculations yield reduced values of ~20–22 J mol$^{-1}$ K$^{-1}$ for specific heat and ~ 28-32 J mol$^{-1}$ K$^{-1}$ for entropy and thermal expansion coefficient of ~3.5 × 10$^{-6}$ K$^{-1}$ recover values closer to experiment. For graphene, experimental data report a room-temperature specific heat of ~20–22 J mol$^{-1}$ K$^{-1}$ (per mole of atoms), vibrational entropy of ~25–33 J mol$^{-1}$ K$^{-1}$, and a small, negative in-plane thermal expansion coefficient of order −(3-7) × 10$^{-6}$ K$^{-1}$ at 300K; The ideal quantum calculations yield reduced values of ~35–40 J mol$^{-1}$ K$^{-1}$ for specific heat and ~ 35-40 J mol$^{-1}$ K$^{-1}$ for entropy and thermal expansion coefficient of ~5 × 10$^{-6}$ K$^{-1}$ recover values closer to experiment at 300K.

## 3.3. Noise and error mitigation techniques

Quantum noise constitutes a central limitation for variational phonon simulations on near-term quantum hardware, particularly for excited-state calculations that require accurate resolution of closely spaced eigenvalues. The quantum noise model employed in this work consists of a composite channel including depolarizing noise, amplitude damping, phase damping, and readout errors. The depolarizing noise is applied after each gate with probability $p_d = 0.01$, amplitude damping is characterized by relaxation probability $p_a = 0.02$, and phase damping by probability $p_\phi = 0.015$. Readout errors are modeled using an assignment matrix obtained from calibration circuits. The values of these parameters are chosen to be consistent with noise levels observed in current superconducting quantum hardware[47].

Figures 9 illustrates the effect of depolarizing gate noise, amplitude error, Phase damping for single qubit gates and readout errors for single qubit and two qubit gates on the quantum-computed phonon dispersions [48]. Readout errors arise from imperfect discrimination of computational basis states during measurement. The errors are captured through classical assignment matrix $M$, where $M_{ij}$ denotes the probability of measuring outcome $j$ when state $i$ is prepared. This model captures the dominant error sources in near-term quantum hardware. Noise perturbs the prepared quantum states, integrate bias and variance in Pauli expectation values, and distorting the optimization landscape of VQE and VQD. In noisy environments, phonon frequencies shows enhanced statistical scatter and systematic shifts from classical reference, and errors arising in high energy optical modes require more complex circuit optimization [49]. These algorithms increase sensitivity to noise due to the inclusion of orthogonality penalty terms, leading mode mixing and loss of symmetry driven features and impact of noise on branched resolved phonon, as depicted in Figures S8-S9, ESI. These findings show that noise-induced errors can mask meaningful vibrational trends. The details are included in the Electronic Supplementary Information (ESI).

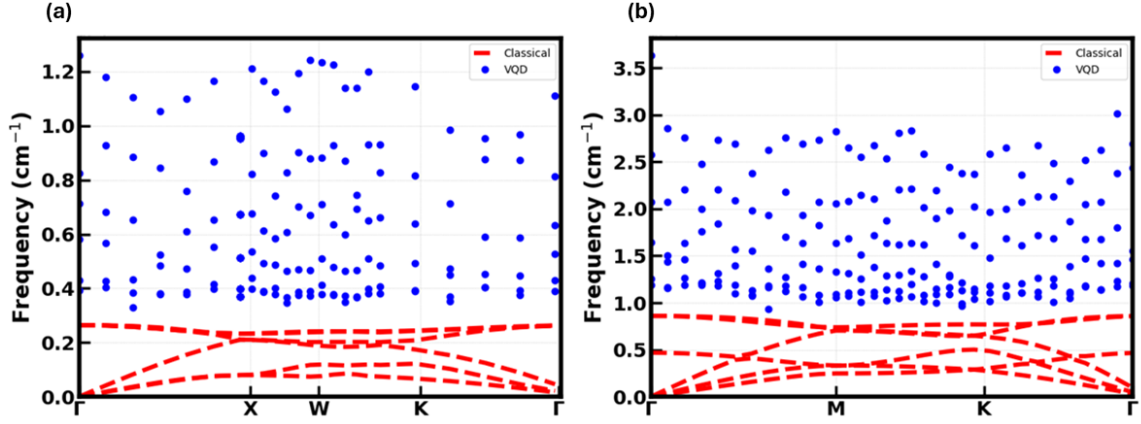

**Figure 9:** Comparison of classical and quantum-computed phonon dispersions in the presence of realistic noise for (a) Silicon (b) Graphene. Gate and readout errors lead to noticeable distortions in the phonon branches.

Error mitigation techniques provide a practical approach to improving the accuracy of variational quantum simulations on near-term quantum hardware. Figure 10 illustrates the noise-induced distortions significantly impact quantum computed phonon dispersions, notably in excited optical modes and branched resolved phonon modes shown in Figure S10-S11, ESI The application of error mitigation techniques—including zero-noise extrapolation [50], readout error mitigation [51], and dynamical decoupling [52, 53] substantially suppresses statistical scatter, restores symmetry-imposed features near the $\Gamma$ point, and improves agreement with classical reference spectra across both acoustic and optical branches. Readout error mitigation is performed by estimating the assignment matrix through calibration circuits and applying its inverse to the measured probabilities. [54].

$$p_{corr} = M^{-1} p_{obs} \tag{10}$$

Here, $\mathbf{p}_{\text{obs}}$ represents the vector of measured probabilities obtained from the quantum device, and $\mathbf{p}_{\text{corr}}$ denotes the corrected probability vector after mitigation of readout errors. The matrix $M$ is the readout error assignment matrix, where each element $M_{ij} = P(\,i \mid j\,)$ represents the probability of measuring the outcome $i$ when the system is prepared in the computational basis state $j$. The inverse matrix $M^{-1}$ is used to reconstruct the ideal probability distribution from the noisy measurements. In practice, $M$ is obtained through calibration experiments by preparing all computational basis states and measuring the corresponding output distributions. Depolarizing noise was modeled as a uniform random Pauli error channel applied after each gate, amplitude damping was included to capture energy relaxation processes, and measurement noise was represented by a classical bit-flip channel. This formulation ensures that the simulated noise reflects the dominant error sources observed in current superconducting qubit platforms as depolarizing noise:

$$\rho \rightarrow (1-p)\rho + p\frac{I}{2^n} \tag{11}$$

and amplitude damping:

$$E_0 = \begin{pmatrix} 1 & 0 \\ 0 & \sqrt{1-\gamma} \end{pmatrix}, E_1 = \begin{pmatrix} 0 & \sqrt{\gamma} \\ 0 & 0 \end{pmatrix} \tag{12}$$

In the above expressions, $\rho$ denotes the density matrix representing the quantum state of the system, and $p$ is the depolarizing error probability, which quantifies the strength of gate noise. The operator $I$ is the identity matrix acting on the Hilbert space of dimension $2^n$, where $n$ is the number of qubits in the system. For the amplitude damping channel, $E_0$ and $E_1$ are the Kraus operators describing energy relaxation processes. The parameter $\gamma$ represents the probability of decay from the excited state $|1\rangle$ to the ground state $|0\rangle$, corresponding to $T_1$ relaxation in quantum hardware. The corrected probabilities are then used to compute expectation values of Pauli operators. In phonon calculations, where Hamiltonians consist of many Pauli terms, readout error mitigation reduces systematic offsets that would otherwise accumulate and distort phonon frequencies. Zero-noise extrapolation reduces gate errors by measuring expectation values [55] at scaled noise levels $\lambda_i = s_i \lambda$, through gate folding or circuit stretching. The zero-noise and extrapolating $\langle O(\lambda_i)\rangle$ to $\lambda = 0$ to the zero-noise limit utilizing exponential fitting. The dynamical decoupling reduces decoherence from low-frequency noise during idle circuits, through control pulses sequences that refocus qubit environment interactions. These approaches reduced statistical scatters, restoration of symmetry in phonon dispersions, and enhanced agreement across both acoustic and optical modes, significant for branch-resolved phonon simulations.

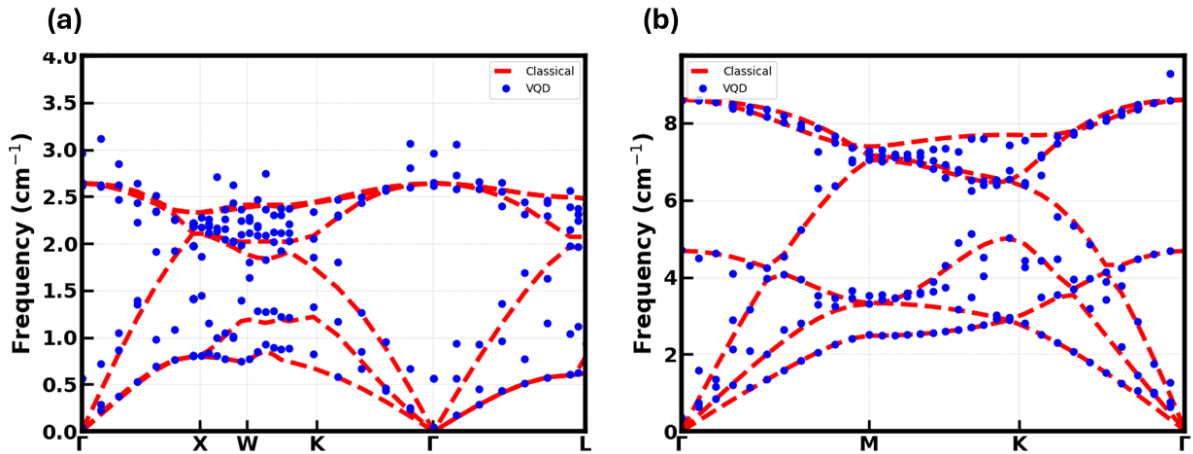

**Figure 10:** Restoration of phonon dispersions across the full high-symmetry path of (a) Silicon (b) Graphene using combined readout mitigation, zero-noise extrapolation, and dynamical decoupling.

The recovery of phonon dispersions arises from these approaches, which complement roles in mitigation measurement bias, gate errors, and decoherence [56]. This mitigation techniques are crucial for examined the feasibility of variational quantum algorithms for computing phonon spectra and thermodynamic characteristics. Without mitigation, noise-induced artifacts obscure physically meaningful trends and prevent reliable comparison with classical lattice-dynamical results. By incorporating error mitigation, deviations between quantum and classical phonon spectra can be attributed to intrinsic algorithmic limitations rather than hardware imperfections. This enables a faithful evaluation of variational quantum algorithms as tools for lattice-dynamical and thermodynamical simulations in the NISQ regime. The quantitative agreement between classical and VQD phonon dispersions is summarized in Table 2 for Silicon and Graphene after error mitigation techniques. Acoustic branches show relative deviations below 10%, while optical modes exhibit slightly larger deviations near the zone boundary.

These results demonstrate that effective error mitigation is essential for obtaining provide a consistent framework for obtaining branch-resolved phonon spectra and derived thermodynamical properties on near-term quantum hardware. The impact of noise and subsequent recovery through error mitigation on thermodynamic quantities is quantified in Table 3, where both entropy and thermal expansion coefficients show significant suppression in the noisy case and partial restoration after mitigation. Further implementation details and quantitative analyses of the mitigation strategies employed in this work is provided in the Electronic Supplementary Information.

**Table 2:** Branch-wise comparison of classical and variational quantum deflation (VQD) phonon frequencies along the Γ–M path for graphene and Γ–X for Silicon, showing root-mean-square error (RMSE), mean absolute error (MAE), and relative deviation after error mitigation techniques.

| Branch | Silicon | | | Graphene | | |
|---|---|---|---|---|---|---|
| | RMSE (cm$^{-1}$) | MAE (cm$^{-1}$) | Relative Error (%) | RMSE (cm$^{-1}$) | MAE (cm$^{-1}$) | Relative Error (%) |
| ZA | 0.22 | 0.18 | 8.0 | 0.18 | 0.14 | 2.5 |
| TA | 0.35 | 0.28 | 9.5 | 0.25 | 0.19 | 3.1 |
| LA | 0.30 | 0.24 | 7.5 | 0.12 | 0.09 | 3.0 |
| ZO | 0.18 | 0.14 | 10.5 | 0.15 | 0.11 | 6.0 |
| TO | 0.25 | 0.20 | 12.0 | 0.17 | 0.13 | 7.2 |
| TO | 0.28 | 0.22 | 13.5 | 0.20 | 0.15 | 8.5 |

**Table 3:** Quantitative comparison of saturation temperatures of $C_V$, vibrational entropy $S$, and thermal expansion coefficient $\alpha$ obtained from classical lattice dynamics and variational quantum simulations under ideal, noisy, and error-mitigated conditions.

| Quantity | Classical | Ideal Quantum | Noisy | Mitigated |
|---|---|---|---|---|
| **C$_v$ saturation temperature (K)** | ~400 K | ~430 K | ~200 K | ~380 K |
| **S (J/mol. K)** | ~90 | ~75 | ~40 | ~70 |
| **α (k$^{-1}$ x 10$^{-6}$)** | ~9 | ~7 | ~5 | ~7 |

This work should be interpreted as a proof-of-principle benchmark demonstrating the applicability of variational quantum algorithms to lattice dynamical problems, rather than as a replacement for established classical methods. The present study is restricted to the harmonic approximation and selected high-symmetry directions of the Brillouin zone. Extension to full Brillouin zone sampling and inclusion of anharmonic effects would require significantly increased computational resources and remains an important direction for future work.

## 4. Conclusions

This work demonstrates that lattice-dynamical calculations provide a physically transparent and stringent benchmark for evaluating variational quantum algorithms beyond electronic structure problems. By mapping first-principles phonon Hamiltonians onto qubit-encoded operators, we show that variational quantum methods can capture phonon branches with reasonable agreement and preserve their relative energy scales. Our results highlight the critical importance of ansatz design and optimization stability in resolving closely spaced vibrational modes, particularly for excited-state calculations that directly impact thermodynamic properties. Using the quantum-computed phonon spectra, we obtained temperature-dependent behavior of specific heat, entropy, and thermal expansion that is consistent with established physical trends, while deviations in absolute values arise from the harmonic approximation and the presence of quantum noise. We further demonstrate that combined error mitigation strategies help recover phonon dispersions and thermodynamic behavior consistent with expected trends on near-term quantum hardware. Although classical phonon methods remain computationally more efficient, the present study establishes phonon-based thermodynamics as a robust and interpretable testbed for assessing variational quantum algorithms and guiding future developments in quantum simulations of lattice-dynamical and thermal properties.

## Acknowledgements

The results presented in this paper has been obtained utilizing the computational facility at Central University of Punjab, Bathinda. NK gratefully acknowledges the financial support during his stay at CQST, SOA University, Bhubaneswar.

## Declaration of competing interests

The authors declare that they have no known competing financial interests or personal relationships that could have appeared to influence the work reported in this paper